\documentclass[5p,times]{elsarticle}

\usepackage{amsmath}
\usepackage{amssymb}
\usepackage{bbm}
\usepackage{graphicx}
\usepackage{subcaption}
\usepackage{multirow}
\usepackage{wrapfig}

\journal{Journal of XYZ}

\begin{document}

\begin{frontmatter}

\title{DeF-DReL: Systematic Deployment of Serverless Functions in Fog and Cloud environments using Deep Reinforcement Learning}

\author[a]{Chinmaya~Kumar~Dehury}
\ead{chinmaya.dehury@ut.ee}
\author[a]{Shivananda~Poojara}
\ead{poojara@ut.ee}
\author[c]{Satish~Narayana~Srirama\corref{cor1}}
\ead{satish.srirama@uohyd.ac.in}
\address[a]{Institute of Computer Science, University of Tartu, Tartu, Estonia}
\address[c]{School of Computer and Information Sciences, University of Hyderabad, Hyderabad 500046, India}
\cortext[cor1]{Corresponding author}

\begin{abstract}
Fog computing is introduced by shifting cloud resources towards the users' proximity to mitigate the limitations possessed by cloud computing. Fog environment made its limited resource available to a large number of users to deploy their serverless applications, composed of several serverless functions. One of the primary intentions behind introducing the fog environment is to fulfil the demand of latency and location-sensitive serverless applications through its limited resources. The recent research mainly focuses on assigning maximum resources to such applications from the fog node and not taking full advantage of the cloud environment. This introduces a negative impact in providing the resources to a maximum number of connected users. To address this issue, in this paper, we investigated the optimum percentage of a user's request that should be fulfilled by fog and cloud. As a result, we proposed DeF-DReL, a Systematic Deployment of Serverless Functions in Fog and Cloud environments using Deep Reinforcement Learning, using several real-life parameters, such as distance and latency of the users from nearby fog node, user's priority, the priority of the serverless applications and their resource demand, etc. The performance of the DeF-DReL algorithm is further compared with recent related algorithms. From the simulation and comparison results, its superiority over other algorithms and its applicability to the real-life scenario can be clearly observed. 
\end{abstract}
\end{frontmatter}


\section{Introduction}\label{sec:intro}
Fog computing becomes an integral part of the cloud computing environment by complementing the practical limitations of the cloud, such as reducing the network overhead, network latency, enabling real-time service, leveraging the QoS by providing location-sensitive services etc \cite{9013785,7439752,NAYERI2021103078}. This is achieved by shifting the computing resource to the end-user's proximity. Fog computing primarily differs from cloud computing in terms of its distance from the users and the resource capacity. The fog environment is generally set up with limited computing and storage resources and can provide service to a limited number of users. Further, very selected small scale services are offered using those limited resources, such as Internet of Thing (IoT) services, smart city services, transportation and logistic services, smart irrigation and agriculture services and many more \cite{9013785,8859632,MAJEED2021103007}. 

\begin{figure}[ht]
    \centering
    \includegraphics[width=0.95\linewidth]{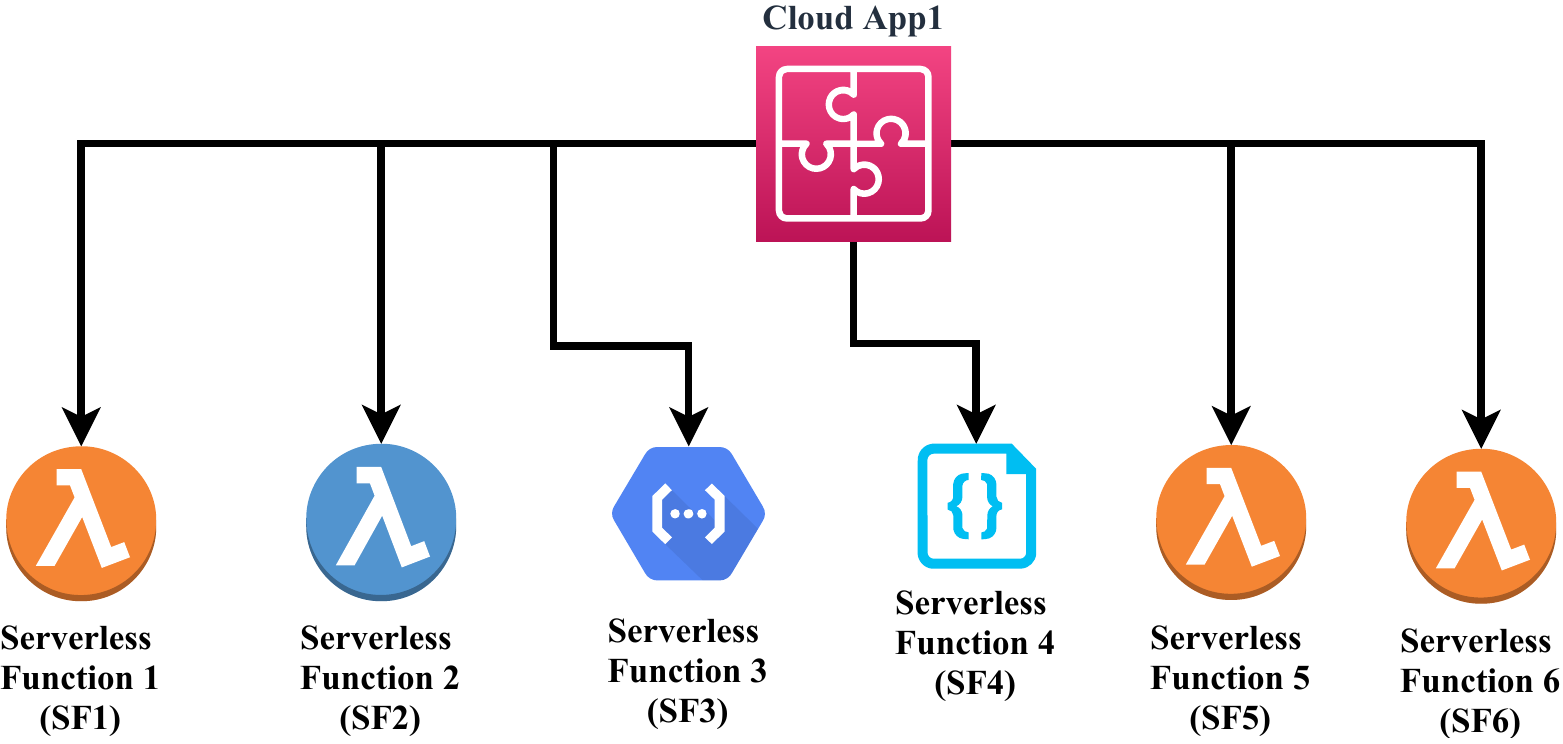}
    \caption{An example of serverless cloud application .}
    \label{fig:ServerlessCloudApp_example}
\end{figure}

The limited fog computing resources are also used to deploy customized users' applications. For efficient and cost-effective utilization of the resources, the fog environment provides a serverless platform that enables a user to deploy a serverless application composed of numerous serverless functions, thanks to container and microservice technology advancement \cite{baldini2017serverless}. Several software solutions, such as OpenFaaS\footnote{https://www.openfaas.com/}, OpenWhisk\footnote{https://openwhisk.apache.org/}, Kubeless\footnote{https://kubeless.io/}, etc., are available in the market to set up a serverless platform. These serverless functions are generally developed for performing small tasks and hence consume fewer resources and can be run within a container  \cite{mohanty2018evaluation}. However, in the case of the higher resource demand of these functions, the fog environment may offload the corresponding serverless functions to the cloud computing environment \cite{ABURUKBA2021102994}. As a result, some portion of the whole applications is deployed in fog, and the other portion is in the cloud environment. For instance, a user sends a request to deploy a serverless application, as shown in Figure \ref{fig:ServerlessCloudApp_example}, which consists of six serverless functions. Upon receiving the application (\textit{Cloud App 1}) by the fog serverless platform, two functions (\textit{SF1} and \textit{SF2}) can be deployed on the fog environment and four functions (\textit{SF2}, \textit{SF3}, \textit{SF4}, and \textit{SF6}) can be deployed on cloud environment, as shown in Figure \ref{fig:motivation_example}.

To efficiently distribute the functions among fog and cloud environments, several approaches are followed, such as the heuristic and AI-based approaches. It is observed that heuristic approaches are mainly difficult when it comes to considering a number of real-life parameters. On the contrary, deep machine learning methods can analyse and process substantial complex problems considering a large number of associated parameters \cite{SHUJA2021103005}. Parameters in this context include resource constraints of serverless platforms at both fog and cloud environment, user's distance, priority, functions resource demand, energy consumption, the latency of the serverless applications, cost of application deployment etc. In this paper, we take advantage of the capability of Deep Reinforcement Learning (DRL), a division of artificial intelligence, while deciding the hosting environment of each serverless function. 

\subsection{Motivation and goal}
It is observed from the recent literature survey that the existing workload offloading solutions mainly focuses on assigning the maximum workload of a particular application to the fog environment \cite{ 8240666,8360511,sami2021demand}. By doing so, the quality of that application is further improved. Here, workload offloading strategy mainly refers to the mechanism to distribute the workload of an application among multiple computing environments. In this work, we consider the deployment of serverless applications atop fog and cloud computing environment. It is assumed that both the computing environments are equipped with the required serverless platform. When the users request the nearby fog computing environment (or the fog node) for their application deployment, the fog node may offload some of the workloads to the cloud computing environment (or cloud node) without violating the SLA (Service Level Agreement) parameters. Assigning more workload to the fog node will significantly reduce the network latency, where assigning more workload to the cloud node significantly reduce the computation latency.  

\begin{figure}[ht]
    \centering
    \includegraphics[width=0.75\columnwidth]{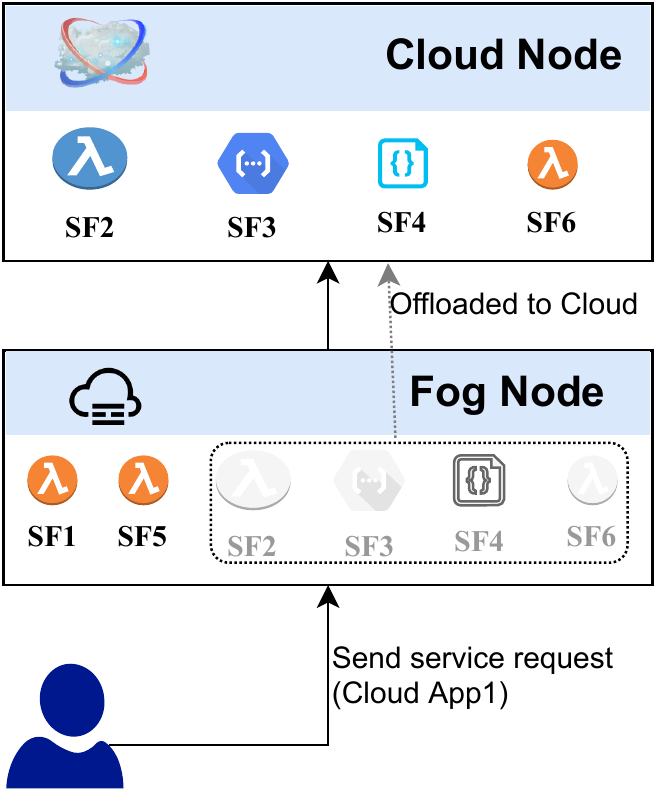}
    \caption{A motivational example of serverless functions deployment in fog and cloud.}
    \label{fig:motivation_example}
\end{figure}

As given in the Figure \ref{fig:motivation_example} example, the serverless function \textit{SF1} and \textit{SF5} of \textit{Cloud App 1} are assigned to the fog node, and the rest of the functions (\textit{SF2}, \textit{SF3}, \textit{SF4}, and \textit{SF6}) are assigned to the cloud node.  It is possible that all six serverless functions are assigned to either the fog node or cloud node only. By doing so, the Cloud App 1 may suffer from larger computation latency or larger network latency, respectively. While deciding the hosting environment of each serverless function, several parameters, such as resource availability of fog node and resource demand of serverless functions, number of connected users, etc., need to be taken into account. This leverages the optimization of function deployment problem more complex. 

Further assigning more number of serverless functions of a particular application to the fog node makes the fog resources unavailable for other serverless applications, which would become a biased solution. Such biased solutions ignore the primary intention of a fog node, which is providing the resources to the applications of the maximum number of connected users at their proximity. This hinted at us to design a novel algorithm that tries to assign the minimum number of serverless functions of an application to the fog node and offload the maximum number of functions to the cloud node without compromising the quality of service or the users' constraints, which is the goal of this paper. By doing so, the number of benefited users can further be maximized. By doing so, the number of benefited users can further be maximized. This motivates us to propose a systematic deployment of serverless functions in fog and cloud environment, as discussed in further sections. 
Towards achieving the goal, our contributions to this paper can be summarized as below:
\begin{itemize}
    \item The problem of efficient deployment of serverless applications on fog and cloud environments is investigated and formulated. 
    \item A novel strategy is designed and developed using the deep reinforcement learning that tries to assign minimum number of serverless functions to fog node. 
    \item A number of real-life parameters are incorporated while developing the agent for the DRL method.  
    \item The developed DRL-based deployment strategy is implemented and compared for the performance evaluation.
\end{itemize}

The rest of the paper is organized as follows: A systematic survey related to the discussed research issue is presented in Section \ref{sec:relWorks}. The mathematical formulation of the concerned serverless application deployment problem is formulated in Section \ref{sec:probForm}, followed by the proposed deployment strategy with the detailed description in Section \ref{sec:Sol}. The detailed implementation and performance results are discussed in Section \ref{sec:perfEval}, followed by the concluding remarks and the scope for improvements in Section \ref{sec:conc}.

\section{Related works} \label{sec:relWorks}
Many research attempts have shown that serverless frameworks deployed on the fog computing environment have several benefits over state of art resource deployment strategies \cite{sarkar2019serverless,grossmann2019applicability,baresi2019towards}. However, serverless function placement over fog or cloud can be decided based on end-user expectations \cite{bermbach2020towards,rausch2021optimized,das2020performance}. Along the side, recent advancements in Artificial Intelligence (AI) algorithms can deliver the desired intelligent decisions \cite{nayeri2021application}. Based on this context, we now describe the literature survey and outline the works relevant to our proposed work.

DRL is being a part of AI, applied in the diverse areas likely in the area of edge and fog computing fields such as resource management \cite{guo2019trusted,qu2020dmro}, service orchestration \cite{guo2021endogenous,chen2020seek}, resource protection \cite{liang2020deep,yang2020resource} and energy-efficient resource management \cite{liu2019intelligent,sami2021demand}. However, the proposed work aligns in a similar direction focusing on optimizing the placement of users' serverless application on fog and cloud environments based on real-life parameters, such as priority of the serverless applications, resource constraints of each serverless platform, distance and latency of the users' from nearby fog node, users' priority, and their resource demand, etc.

Bermbach et al. \cite{9103477}, proposed auction-based function placement for optimal decision to execute serverless functions on edge and cloud to maximize the revenue considering the execution cost and arrival rate of users' requests. The major objective in \cite{9103477} is to reduce cost without considering the latency and priority needs of the user applications and resource demands.  Similarly, Cheng et al. \cite{cheng2019fog} proposed a data-centric programming model called Fog Function for placement and scaling of serverless functions based on data context, system context, and usage context considering latency and flexibility factors. This work is focused on latency and data-oriented requirements. However, it lacks in considering nearby fog nodes and the priority of users' requests.


Serverless function placement resembles the problem of service placement in fog and clouds. Several attempts have been made to address this problem using heuristic and machine learning-based approaches. Hamed et al. \cite{8360511} proposed a game theory to allocate the resource to IoT users' in hierarchical computing paradigm, including fog and remote cloud computing services. They formulated a computation offloading game to model the competition between IoT users' and allocate the limited processing power of fog nodes efficiently. The concerned research problem is purely modelled with the game theory-based heuristic approach, which might be less sustainable in mobile and stochastic IoT environments.  Similarly, Du et al. \cite{8240666} proposed a computation offloading problem in a mixed fog/cloud system by jointly optimizing the offloading decisions and the allocation of computation resource, transmit power, and radio bandwidth while guaranteeing user fairness and maximum tolerable delay.

In our previous work \cite{dehury_ccgrid}, a novel DRL mechanism for optimized service delivery by slicing the user requests is proposed that distribute the workload among fog and cloud without comprising the service quality. The major limitation of the proposed mechanism lies in its inability to take the user's priority and its inapplicability to the modern serverless platform. A deep reinforcement learning method is used in Chen et al. \cite{CHEN20211} to improve the performance of mobile edge computing. The problem of offloading is modelled using Markov Decision Problem (MDP) and offloads the workload to cloud computing if the mobile edge devices do not possess enough resources to handle the workload.  The  Deep  Sequential Model-Based on Reinforcement Learning method is used by Wang et al. \cite{wang2019computation} to decide whether to offload the workload to cloud computing. The latency, task dependency and the communication topology are taken into consideration in the proposed mechanism. However, the proposed strategy may not be able to take advantage of the available resources at the nearby fog node, which could have further reduce the latency. 

To the best of our knowledge and from the literature survey, it is observed that the related recent works focus on using game theory, auction-based approach, heuristic approaches etc. for strategic deployment of users' applications considering a limited number of parameters making them not enough matured and hard to apply in real life. To fill this gap between the scientific research and its practical applicability, we propose a novel DRL-based systematic deployment of users application on both nearby fog and cloud environment. Unlike the existing works, the proposed strategy considers the number of users connected to the fog node, their distance and network latency,  serverless computing and the associated constraints, such as short-run function execution, limited resource demand, environment-specific resource constraints, etc. 


\section{Problem formulation}\label{sec:probForm}
With the above motivation and the goal we envisioned to achieve, in this paper, we have considered an environment where many users send their resource requests to a nearby fog node. The fog node is further backed by the hypothetically infinite amount of resources of the cloud environment. However, the number of fog nodes and cloud nodes can be increased to fit into the desired situation. The fog node is responsible for providing service to all the connected users with the intention to fulfil their requirements. The user's request can be interpreted as the deployment of their serverless applications on the fog node. The serverless application is composed of several serverless functions that can be deployed on either the fog environment or cloud environment. In the following subsections, the detailed modelling of deployment environments, the users, and their requests are presented. For quick reference, we have included a list of all the notations used in this manuscript in chronological order in Table-\ref{table:Notation}.


\begin{table}
\caption{List of Notation.}
\begin{tabular}{|p{0.15\linewidth}|p{0.75\linewidth}|}
    \hline
    \textbf{Notation} & \textbf{Description} \\ \hline
    $\dot{A}$ & The action space.  \\ \hline
    $C$ & cloud environment \\ \hline
    $\mathbbm{c}^i_j$ & Computation latency of the serverless function $\hat{s}^i_j$ \\ \hline
    $ \mathbbm{C}_i $ & Computation latency of the SSR $s_i$ \\ \hline
    $d_i$ & User's distance from the Fog node \\ \hline
    $D$ & Maximum distance a user is allowed from Fog node \\ \hline
    $F$ & The fog node \\ \hline
    $l^f$ & Communication Latency between fog and cloud \\ \hline
    $l_i$ & Communication Latency between fog and user \\ \hline
    $\hat{l}^i_j$ &  Communication latency of serverless function $\hat{s}^i_j$\\ \hline
    $n$ & Total number of users connected to the fog node \\ \hline
    $P_i$ & Priority of User $u_i$ \\ \hline
    $p^{\omega}$ & Importance factor of latency and distance.\\ \hline
    $\mathbbm{P}_j^i$ & Priority of serverless function $\hat{s}^i_j$ \\ \hline
    $K_j^i$ & Size of the serverless function code $\hat{s}^i_j$. \\ \hline
    $\hat{K}^f $ & Function code size limit in the fog’s serverless platform.\\ \hline
    $\hat{K}^c $ & Function code size limit in the cloud’s serverless platform.\\ \hline
    $ \kappa_j^i $ & Represent the size of the INPUT to the function $\hat{s}^i_j$. \\ \hline
    $ \hat{\kappa}^f $ & The input limit to any function in the fog’s serverless platform.\\ \hline
    $ \hat{\kappa}^c $ & The input limit to any function in the cloud’s serverless platform.\\ \hline
    
    $\hat{R}^f(x)$ & Upper limit of resource demand of serverless function in FOG \\ \hline
    $\hat{R}^c(x)$ & Upper limit of resource demand of serverless function in Cloud \\ \hline
    $\hat{r}^i_j(x)$ & Upper limit resource demand of a serverless function $\hat{s}^i_j$ \\ \hline
    $r^i_j(x)$ & Total resource demand of a serverless function $\hat{s}^i_j$ \\ \hline
    $S$ & SSR Bucket or set of SSRs from at most n number of users; maximum of one from each user  \\ \hline
    $s_i$ & SSR from the user $u_i$  \\ \hline
    $s_j^i$ & Serverless function belong to SSR $s_i$  \\ \hline
\end{tabular}\label{table:Notation}
\end{table}

\subsection{Deployment environments: Fog and Cloud}
Let $F$ be the Fog Node (FN) equipped with a limited amount of resources. The resource here refers to CPU, RAM, Storage, and Network I/O. The FN is responsible for providing services/resources to a maximum number of connected users. The cloud node, denoted by $C$, is equipped with a large number of interconnected high-end servers. The FN is responsible for receiving the request from the users to deploy their serverless application that composed of a number of serverless functions. Upon receiving the request, the fog node decides the deployment environment of each function. In such a decision process, the communication latency, denoted by $l^f$, is considered as one of the major parameters. 

\subsection{Modelling of Users}
Let $n$ be the total number of users connected to the FN. $U = \{ u_1, u_2, \dots, u_n\}$ represents the set of all connected users. Each user is associated with two major parameters: (a) geographic distance from the FN and (b) the communication latency from the FN. $(x, y)$ and $(p_i, q_i)$ denotes the location of FN $F$ and user $u_i$, respectively. The shortest possible path through space between fog and the user can be calculated using the Euclidean distance method, as follows:

\begin{equation}\label{eq:distance}
    d_i = \sqrt{(p_i - x)^2 + (q_i - y)^2}
\end{equation}

Let $D$ be the maximum distance that a user can access the service from the nearby fog node. Based on $D$ value and the current distance of the user from the FN, the distance-based priority can be calculated as follows:
\begin{equation}\label{eq:priority_dist}
    P_i^d = \frac{d_i}{D}
\end{equation}

It can be seen that the value of $P_i^d$ always lies between $0$ and $1$(inclusive). A higher value of $P_i^d$ indicates the longer user's distance from FN.
Like the distance, the communication latency plays a major role in the deployment environment selection process for each serverless function. The communication latency is nothing but the Round Trip Delay (RTD) for a single data packet between the user and the FN. 
It is evident that latency increases when the distance between the user and FN increases. However, this may not hold due to the local environmental characteristics and physical obstructions and hence the users with a shorter distance from the FN may not experience lower latency \cite{tse2005fundamentals}.
Let $l_i$ be the latency for the user $u_i$. Based on the latency, the priority of a user $u_i$ can be calculated as

\begin{equation}\label{eq:priority_latency}
    P_i^l = \frac{l_i}{max\{l_j\}}, \forall u_j
\end{equation}

By taking the distance-based priority, as in Equation \ref{eq:priority_dist} and the latency-based priority, as in Equation \ref{eq:priority_latency},   into account, the overall priority, $P_i$, of the user $u_i$, can be calculated as below.
\begin{equation}\label{eq:priority_user}
    P_i = [p^\omega * P_i^d] + [P_i^l * (1-p^\omega)]
\end{equation}

where, $p^{\omega}$ represents the importance factor of latency and the distance-based priority. $p^{\omega}$ is useful for the fog environment to give priority to the latency parameter and distance. If $p^{\omega}$ is high, the fog environment gives higher priority to the “distance” parameter and will have negligible importance to the “latency” parameter. In other words, users will be assigned with the priority value based on their distance from the fog node. This is useful for the service provider when more number of users are far from the fog node (towards the edge of the communication range of the fog node). Further, the service requests from the users with high priority, i.e. high $P_i$ value, will be preferably handled by the fog environment instead of the cloud environment. 
It is assumed that the value of $p^{\omega}$ will be determined by the service provider. From the above Equation \ref{eq:priority_user}, it can be concluded that the value of $P_i$ is dynamic and depends on user's current location and latency and will be the part of SLA.  

\subsection{Serverless Service Request (SSR)}
As discussed before, users send their requests to the FN to deploy their serverless applications. Each such request is called Serverless Service Request (SSR). Let $S = \{ s_i, s_2, \dots, s_n\}$ be the set of such SSRs, also known as \textit{SSR bucket}. The SSR bucket may contain a maximum of $n$ number of requests, and each user can send a maximum of one SSR. The term $s_i$ represents a SSR from user $u_i$. Further, each SSR $s_i$ is nothing but a serverless cloud application that is to be deployed on FN and cloud environment. The SSR $s_i$ is composed of a set of serverless functions and can be defined as $s_i = \{ \hat{s}_1^i, \hat{s}_2^i, \hat{s}_3^i, \dots \}$. It is also possible that the set $s_i$ can be an empty set representing no request from the user $u_i$. All the SSRs may composed of different number of serverless functions, i.e. $|s_i| \ne |s_k|, i\ne k, \forall s_i, s_k \in S$. 

\begin{figure}[ht]
    \centering
    \includegraphics[width=0.95\linewidth]{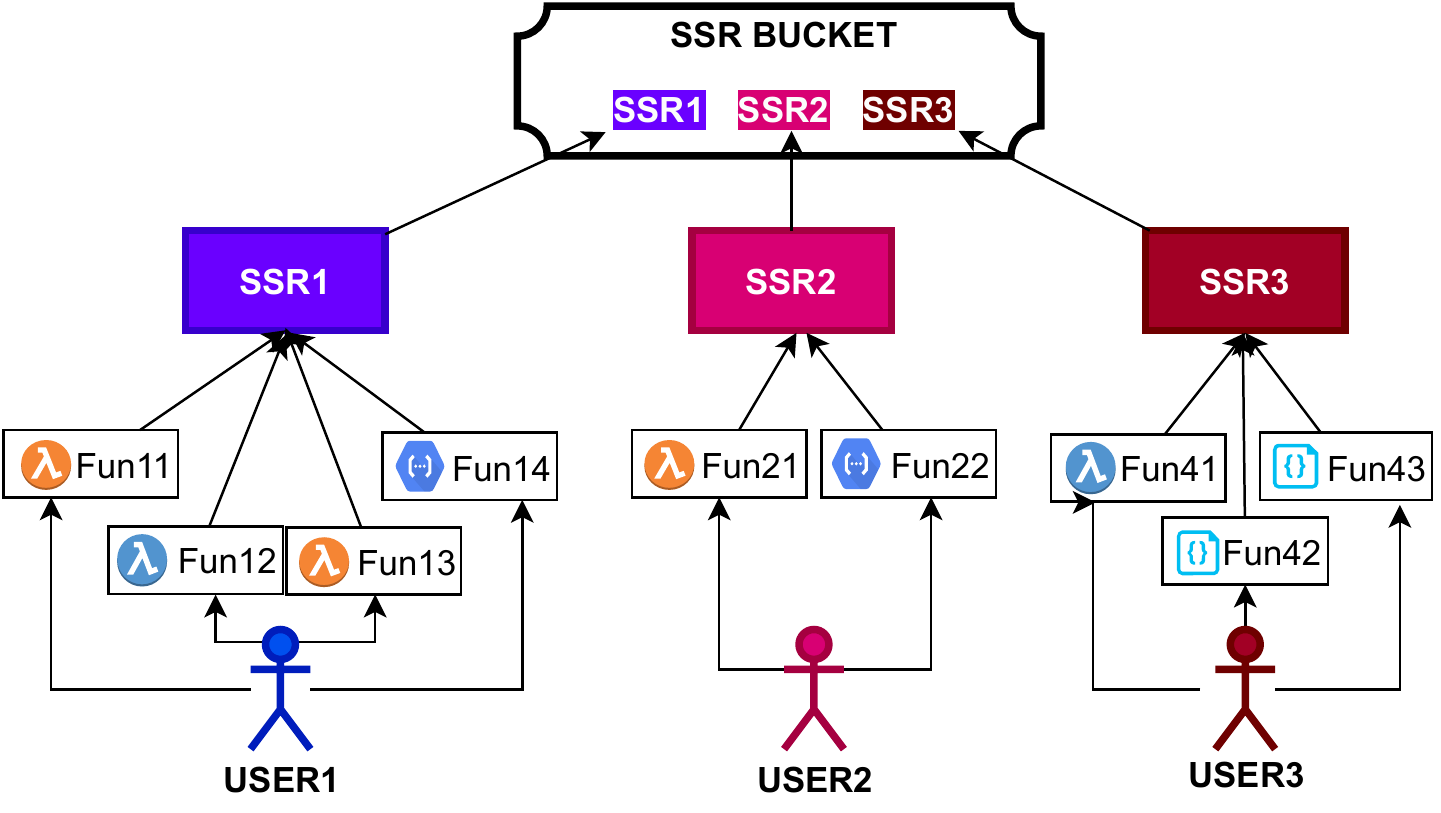}
    \caption{An example of SSR bucket.}
    \label{fig:SSRbucket_example}
\end{figure}

The example in Figure \ref{fig:SSRbucket_example} further illustrates the SSR and SSR bucket. In the example, the SSR bucket contains the three SSRs ($SSR1, SSR2,$ and $SSR3$) from users $USER1, USER2,$ and $USER3$, respectively. $SSR1$ serverless cloud application from $USER1$ is composed of four serverless functions ($Fun11$, $Fun12$, $Fun13,$ and $Fun14$). Similarly, the SSRs from $USER2$ and $USER3$ are composed of two and three serverless functions, respectively. As a whole, upon receiving the SSR bucket, the FN needs to decide the hosting serverless platform (in fog or in the cloud) of all the nine serverless functions.

The boolean variable $f(\hat{s}_j^i)$, where $f(\hat{s}_j^i)=1$, indicates that the serverless function $\hat{s}_j^i$ is deployed on FN, else deployed on cloud.
Similarly, the counterpart boolean variable $c(\hat{s}_j^i)$ indicates if the function $\hat{s}_j^i$ is deployed on cloud environment. 
Both the boolean variables are counterpart to each other and must follow condition $f(\hat{s}^i_j) + c(\hat{s}^i_j) = 1$.

\subsubsection{Resource demand}
The functions are deployed on the either fog or cloud serverless platforms as discussed earlier. Each serverless platform posses different resource limits. In other words, each serverless functions have the limit on use of different resource types. It is assumed that, the resource limit in fog serverless platform is less than that of in the cloud serverless platform. Let $\hat{R}^f(x)$ and $\hat{R}^c(x)$ be the upper limit of type $x$ resource consumption for any serverless function in fog and cloud environment, respectively, where $x \in \{ CPU, RAM, Disk, Network I/O \}$. The value of $\hat{R}^f(x)$ and $\hat{R}^c(x)$ are decided by serverless platform service provider. For any serverless function $\hat{s}^i_j \in s_i$, the maximum amount of type $x$ resource demand can be defined as 
\begin{equation}\label{eq:resrcLimit_function}
    \hat{r}^i_j(x) = \left[f(\hat{s}^i_j) * \hat{R}^f(x)\right] + \left[c(\hat{s}^i_j) * \hat{R}^c(x)\right]
\end{equation}
The value of $\hat{r}^i_j(x)$ depends on its deployment environment. If the serverless function is deployed on fog environment, the maximum amount of type $x$ resource it can demand would be $\hat{R}^f(x)$, $\hat{R}^c(x)$ otherwise.

However, to maintain the workload balance on different resources, the Resource Importance Factor, denoted by $ r^{\omega}(x) $, is introduced to handle the imbalanced resource availability in both environments. Imbalanced resource availability refers to the nonuniform percentage of resource availability in a particular computing environment. 
It is assumed that the value of $r^{\omega}(x)$ will be decided by the service provider and may need to update on a timely basis, based on resource availability. 

The resource demand is not fixed in today's serverless computing. A serverless function may need more resource on the fly. The resource requirement of a serverless function is of two types: (a) Base resource requirement, denoted by $\check{r}^i_j(x)$ and (b) supplementary resource requirement, denoted by $\bar{r}^i_j(x)$. The \textit{base resource demand} refers to the minimum demand of type $x$ resource by the function $\hat{s}^i_j$. Each serverless function may process a variable amount of data given as the input. A serverless function may demand additional resource, also known as a \textit{supplementary resource}, to process that input data. Upon combining the base and the supplementary resource demand, the total resource demand of a serverless function can be calculated as below.
\begin{equation}\label{eq:tot_resrc_req_fun}
     r^i_j(x) = \left[ \check{r}^i_j(x) + \bar{r}^i_j(x) \right] \le \hat{r}^i_j(x), \forall x
\end{equation}

The total base resource requirement of a SSR $s_i$ can be calculated as the sum of base resource requirement of all serverless functions. Mathematically,
\begin{equation}\label{eq:resrc_req_base_ssr}
    \check{R}_i(x) = \sum_{\forall \hat{s}^i_j \in s_i} \check{r}^i_j(x), \quad \forall x
\end{equation}
Similarly, the total supplementary resource requirement of the SSR $s_i$ can be calculated as below.
\begin{equation}\label{eq:resrc_req_supplementary_ssr}
    \bar{R}_i(x) = \sum_{\forall \hat{s}^i_j \in s_i} \bar{r}^i_j(x), \quad \forall x
\end{equation}
The total resource requirement of a SSR can be calculated either by summing up the total resource requirements of all serverless functions or by adding the value in Equation \ref{eq:resrc_req_base_ssr} and \ref{eq:resrc_req_supplementary_ssr}. As discussed before that, each serverless function is designed with a dedicated functionality having a major or minor contribution to the whole serverless application.

\subsubsection{Priority of SSR}
Based on the functionality, users can decide how critical each function is, denoted by $\hat{\mathbbm{q}}^i_j$. Eventually, this critical value of the serverless functions will also play a major role in deciding the deployment environment. The value of $\hat{\mathbbm{q}}^i_j$ ranges from $1$ through $5$.
Function with critical value $5$ can be interpreted as a high priority function and must be deployed on the serverless platform of nearby FN, provided the function's resource demand is within the limit. On the contrary, a function with a critical value $1$ should be deployed on the serverless platform in the cloud environment. Similar to the criticality of a function, $\mathbbm{q}_i$ denotes the criticality of the SSR $s_i$.

Serverless platforms are most suitable for the short run and small-sized functions. In other words, it is not recommended to deploy very large functions that take hours to finish their intended execution. To fit into such a scenario, the size of the serverless functions and the size of their input are incorporated.
The size of the serverless function refers to the size of the code and the dependent library files. On the other hand, input size mainly refers to the size of the data passed to the function as an argument.
$K_j^i$ and $\kappa_j^i$ denotes the size of the serverless function $\hat{s}^i_j$ and its input size, respectively. Input size represents the approximate amount of data passed to the function as an argument. On the other hand, the serverless platform setup in fog and cloud environment possess their own limitations. Let $\hat{K}^f$ and $\hat{K}^c$ denote the limit on function code size in fog and in the cloud environment, respectively. Similarly, $\hat{\kappa}^f$ and $\hat{\kappa}^c$ denote the limit on input size to the serverless functions in fog and in cloud environments, respectively. Smaller function size and functions having smaller input size have a higher preference to be deployed in the fog environment. The priority of a serverless function, denoted by $\mathbbm{P}^i_j$, is derived from it’s size ($K_j^i$), it’s input size ($\kappa_j^i$), and user-specified critical value ($\hat{\mathbbm{q}}^i_j$). Mathematically, 
\begin{align}\label{eq:priority_fun}
    \mathbbm{P}^i_j = 1 \bigg/ \left( \Delta + \left[ \frac{5-\hat{\mathbbm{q}}^i_j}{5} \right] * \left[ \frac{\max_{\forall \hat{s}^i_j \in s_i}\{K_j^i\} - K_j^i }{\max_{\forall \hat{s}^i_j \in s_i}\{K_j^i\}} \right] \right. \nonumber \\ 
    \left. * \left[ \frac{\max_{\forall \hat{s}^i_j \in s_i}\{\kappa_j^i\} - \kappa_j^i }{\max_{\forall \hat{s}^i_j \in s_i}\{\kappa_j^i\}} \right]\right) 
\end{align}
, where $\Delta$ is a small fractional constant, used to avoid \textit{divide by zero} situation. 
The terms in the denominator are normalised to the range between 0 and 1. This is primarily to handle the different units of function size ($K_j^i$), input size ($\kappa_j^i$), and user-specified critical value ($\hat{\mathbbm{q}}^i_j$) parameters. 
Higher the priority value, higher preference for the fog node. It is to be noted that the functions having the same characteristics and are from different SSRs may have different priority value, as it depends on the characteristics of other functions in the same SSR.  

\subsubsection{Communication latency of SSR}
Similar to the user's latency, each serverless function posses a communication latency denoted by $\hat{l}^i_j$, which is defined by the summing up the user's latency and distance, as given in Equation \ref{eq:priority_user} and the communication latency between fog and cloud environment (in case the function is deployed in cloud). Mathematically, 
\begin{equation}\label{eq:commLtncy_fun}
    \hat{l}^i_j = P_i + c(\hat{s}^i_j)*l^f
\end{equation}
Further, by summing up the communication latency of all serverless, we can calculate the communication latency of a SSR $s_i$, as given in Equation \ref{eq:commLtncy_ssr}.
\begin{equation}\label{eq:commLtncy_ssr}
    \mathbbm{L}_i = \sum_{\forall \hat{s}^i_j \in s_i} \hat{l}^i_j
\end{equation}

\subsubsection{Computation latency of SSR}
Similar to the communication latency, each serverless function also posses a computation latency, which depends on the amount of total resource demand, the upper limit of the resource allocation set by the serverless function and the communication latency between fog and cloud environment. Mathematically, 
\begin{align}\label{eq:comp_ltncy_func}
    \mathbbm{c}^i_j = \left[ \sum_{\forall x \in \{CPU, RAM, Storage\}} \frac{r^i_j(x)}{\hat{r}^i_j(x)}*r^\omega(x) \right] + \nonumber \\
    \left[ \frac{r^i_j(I/O)}{\hat{r}^i_j(I/O)} * r^\omega(I/O) * \left[ f(\hat{s}^i_j)*l_i + c(\hat{s}^i_j)*l^f \right] \right]
\end{align}
The average computation latency of a SSR $s_i$ can be calculated by taking the computation latency of all serverless functions, as given in Equation \ref{eq:comp_ltncy_ssr}.
\begin{equation}\label{eq:comp_ltncy_ssr}
    \mathbbm{C}_i = \sum_{\forall \hat{s}^i_j \in s_i } \mathbbm{c}^i_j * \frac{\hat{\mathbbm{P}}^i_j}{\max_{\forall \hat{s}^i_j \in s_i}\{\hat{\mathbbm{P}}^i_j\}}
\end{equation}

\subsection{Objective function}\label{sec:objFun}
Considering the motivation and system model of deployment environments, users, and serverless applications, this research work aims to fulfil the purpose of the fog computing environment by providing the service to a maximum number of users. The primary purpose of fog computing's introduction is to provide real-time, geographic data-sensitive service, and many more, as discussed in Section \ref{sec:intro}. For this, it is necessary to accommodate a maximum number of connected user, which can be achieved by deploying only the required service and offload the unnecessary services to the cloud. Such a problem can be mathematically formulated as follows.

\textbf{Objective} function for SSR $s_i$:
\begin{equation}\label{eq:obj:ssr}
    Minimize ~ z_i = \mathbbm{L}_i + \mathbbm{C}_i 
\end{equation}
The Equation \ref{eq:obj:ssr} indicates the objective while processing the request $s_i$ from the user $u_i$. The objective here is to minimize both the communication latency, as defined in Equation \ref{eq:commLtncy_ssr} and computation latency, as defined in Equation \ref{eq:comp_ltncy_ssr}. However, at any particular time, fog environment receives a set of requests from the connected users, which is denoted as the SSR bucket $S$. Hence it is essential to formulate such a multi-objective optimization (MOO) problem for the whole SSR bucket $S$ and accordingly can be written in following manner (\cite{moo}) in Equation \ref{eq:obj:ssrbucket}.

\textbf{Objective} function for the SSR bucket $S$:
\begin{align}\label{eq:obj:ssrbucket}
    \min z_i, 1 \le i \le n
\end{align}

\textbf{Constraints:}
\begin{equation}\label{const:nonEmptySSR}
    |s_i| > 0, \quad 1 \le i \le n
\end{equation}
\begin{equation}\label{const:funSize_fog}
    if\text{ }f(\hat{s}^i_j)=1, K_j^i \le \hat{K}^f,  \forall \hat{s}^i_j \in s_i\text{ and } \forall s_i \in S
\end{equation}
\begin{equation}\label{const:funSize_cloud}
    if\text{ }c(\hat{s}^i_j)=1, K_j^i \le \hat{K}^c,  \forall \hat{s}^i_j \in s_i\text{ and } \forall s_i \in S
\end{equation}
\begin{equation}\label{const:funInputSize_fog}
    if\text{ }f(\hat{s}^i_j)=1, \kappa_j^i \le \hat{\kappa}^f,  \forall \hat{s}^i_j \in s_i\text{ and } \forall s_i \in S
\end{equation}
\begin{equation}\label{const:funInputSize_cloud}
    if\text{ }c(\hat{s}^i_j)=1, \kappa_j^i \le \hat{\kappa}^c,  \forall \hat{s}^i_j \in s_i\text{ and } \forall s_i \in S
\end{equation}
\begin{equation}\label{const:resrcDmndLimit_fog}
    if\text{ }f(\hat{s}^i_j)=1, r_j^i(x) \le \hat{R}^f(x), \forall \hat{s}^i_j \in s_i\text{ and } \forall s_i \in S
\end{equation}
\begin{equation}\label{const:resrcDmndLimit_cloud}
    if\text{ }c(\hat{s}^i_j)=1, r_j^i(x) \le \hat{R}^c(x), \forall \hat{s}^i_j \in s_i\text{ and } \forall s_i \in S 
\end{equation}
\begin{equation}\label{const:resourceImpFactor}
    \sum_{\forall x}r^{\omega}(x) = 1
\end{equation}

The constraints for above objectives function can be summarized as below:
\begin{itemize}
    \item Constraint \ref{const:nonEmptySSR} ensures that the SSR bucket contains the SSRs that are composed of at least one serverless function.
    \item Constraint \ref{const:funSize_fog} and \ref{const:funSize_cloud} ensure the size of all the serverless functions are not more than the function size limit of the corresponding hosting serverless platform.
    \item Similarly, Constraint \ref{const:funInputSize_fog} and \ref{const:funInputSize_cloud} ensure the size of input to all the serverless functions are not more than the limit of the corresponding hosting serverless platform.
    \item Constraint \ref{const:resrcDmndLimit_fog} and \ref{const:resrcDmndLimit_cloud} ensure that the resource demand of the serverless functions are less than the resource limit imposed by the fog and cloud environments, respectively.
    \item Constraint \ref{const:resourceImpFactor} ensures that the sum of resource importance factor for all different types of resource is 1.
\end{itemize}

\section{Proposed solution: DeF-DReL} \label{sec:Sol}
The previous section discusses the serverless function distribution research challenge. In a nutshell, when a user sends a serverless-based application to the nearby FN, what percentage of the total workload (or the total number of serverless function) should be handled by the fog node and cloud node. In other words, which serverless functions should be deployed on FN and should be deployed on the cloud environment. As discussed, it is essential to deploy a minimum number of functions on FN that minimizes the communication and computation latency without compromising the QoS parameters. Keeping this objective in mind, we have proposed a Deep Reinforcement Learning-based serverless function deployment strategy for fog and cloud computing environments, named as \textit{DeF-DReL}. 

In this section, we will be mainly focusing on the working principle of the proposed deployment strategy that uses Deep Reinforcement Learning (DRL) as a tool in the assessment of which deployment environment would be best suited for each serverless function. A similar approach is used in our previous work \cite{dehury_ccgrid} that focuses on an efficient service dispersal mechanism for both fog and cloud environment using DRL. 
In the DRL approach, a branch of ML and based on Artificial Neural Network (ANN), the features are extracted in a layer-wise fashion unlike the large matrix of state-action pairs, in conventional Q-learning algorithm. This allows us the obtain finer result in a comparatively large-scale simulation/experimental environment \cite{dehury_ccgrid}.
As discussed in Section \ref{sec:intro}, the fundamental components are the state space of the agent, action space that the agent will follow to move forward and all possible rewards for each possible action.  
\subsection{State space}
The state-space consists of all possible combinations of the states of all serverless functions. The state of a serverless function indicates the environment where it is hosted. Mathematically, 
\begin{equation}
    \dot{s}_j^i = < f(\hat{s}^i_j), c(\hat{s}^i_j) > 
\end{equation}
, where $\dot{s}_j^i$ is the state of a serverless function. For instance, $\dot{s}_5^2 = < 1, 0 >$ indicates that the serverless function $\hat{s}^5_2$ is hosted on fog computing environment, which is the state of the this serverless function. However, the state of a serverless function can be $\dot{s}_j^i = < 0, 0 >$, which represents that the function is not assigned to any of the computing environment. By combining the states of all the serverless functions, the state of a SSR $s_i$, denoted by  $\dot{s}_i$ can be formulated as $\dot{s}_i = < \dot{s}_1^i, \dot{s}_2^i, \dot{s}_3^i, \dots >$. Further, the state of the whole SSR bucket $S$ can be represented as $\dot{S} = < \dot{s}_1, \dot{s}_2, \dots \dot{s}_n>$.

\subsection{Action space}
The action of an agent represents its movement from one state to the other state of the SSR bucket. In other words, an action represents the assignment of a serverless function either onto the fog environment or onto the cloud environment. For the agent and SSR bucket $S$, the total action space, denoted as $\dot{A}$ is composed of all the possible movements. The possible actions for an agent are (a) $\dot{a}^f$ representing the actions where the function is assigned to FN and (b) $\dot{a}^c$, representing the actions where the function is assigned to cloud computing environment. 

\subsection{Rewards}
When the agent transitions from one state to another by taking a particular action, its main objective is to optimize the reward. The one-step reward can be broken down to reward on assigning each serverless function present in the SSR bucket. It is to be noted that the reward gained depends on the action taken by the agent. For each action applied to the whole SSR bucket by the agent, the reward gained can be calculated as the average reward gained by applying the same action to all the SSRs. Mathematically,

\begin{equation}\label{eq:reward_SSRBucket}
    \dot{S} = \left( \frac{1}{|S|}\right)\sum_{\forall s_i \in S}\dot{r}(s_i)
\end{equation}
, where $\dot{r}(s_i)$ represents the total reward gained by imposing an action to all the serverless functions present in the SSR $s_i$. This can be calculated as below,
\begin{equation}\label{eq:reward:SSR}
    \dot{r}(s_i) = \sum_{\forall \hat{s}^i_j \in s_i} \left[ \{ f(\hat{s}^i_j) * \dot{r}^f(\hat{s}^i_j)\} + \{ c(\hat{s}^i_j) * \dot{r}^c(\hat{s}^i_j) \} \right]
\end{equation}
, where $\dot{r}^f(\hat{s}^i_j)$ and $\dot{r}^c(\hat{s}^i_j)$ are the rewards gained by the agent by imposing the action $\dot{a}^f$ and $\dot{a}^c$ on the serverless function $\hat{s}^i_j$, as calculated in Equation \ref{eq:reward:fun_on_FN} and \ref{eq:reward:fun_on_cloud}, respectively. 
When a serverless function is assigned to fog node, i.e. if the action $\dot{a}^f$ is imposed on a function $\hat{s}^i_j$, the reward value is in the form of both computation and communication latency. All together the total reward gained by applying action $\dot{a}^f$ on the serverless function $\hat{s}^i_j$ is calculated as

\begin{equation}\label{eq:reward:fun_on_FN}
\begin{split}
    \dot{r}^f(\hat{s}^i_j) & = \left[ \sum_{\forall x \in \{CPU, RAM, Storage\}} \frac{r_j^i(x)}{\hat{R}^f(x)}*r^\omega(x) \right]  \\
    & + \left[ \frac{r_j^i(I/O)}{\hat{R}^f(I/O)}*r^\omega(I/O)*l_i \right] + P_i
\end{split}
\end{equation}
, where $P_i$ is the rewards in the form of communication latency and the rest of the terms in above Equation \ref{eq:reward:fun_on_FN} are the rewards in the form of computation latency. 
Similarly, the total reward that can be gained by imposing the action on $\dot{a}^c$ on the serverless function $\hat{s}^i_j$ is calculated as 
\begin{equation}\label{eq:reward:fun_on_cloud}
\begin{split}
    \dot{r}^c(\hat{s}^i_j) & =  \left[ \sum_{\forall x \in \{CPU, RAM, Storage\}} \frac{r_j^i(x)}{\hat{R}^c(x)}*r^\omega(x) \right] \\  
    & + \left[ \frac{r_j^i(I/O)}{\hat{R}^c(I/O)}*r^\omega(I/O)*\{l_i+l^f\} \right] + P_i + l^f
\end{split}
\end{equation}

\section{Performance evaluation}\label{sec:perfEval}
The proposed DeF-DReL algorithm is evaluated with small scale simulation. This section mainly devoted to providing detailed information on the simulation setup/environment, performance evaluation matrix, simulation results and discussions. The DeF-DReL algorithm deploys the serverless applications on both fog and cloud's serverless platform by assessing several properties and requirements of the serverless applications. For comparison purpose, Computation Offloading Game (COG)~\cite{8360511} and Computation Offloading and Resource Allocation (CORA)~\cite{8240666} algorithms are used. Shah-Mansouri et al. \cite{8360511} study the problem of resource allocation of fog and cloud computing to the connected IoT users and proposed a computation offloading game (COG) algorithm. The IoT devices offload the tasks to the connected fog nodes and the cloud node with the goal to maximize the quality of experience. Similarly, Jianbo et al. in \cite{8240666} investigated the computation offloading problem and proposed an algorithm, named CORA, that considers a similar environment where user equipment offloads the request with the tasks’s metadata. The tasks can be executed on the local device along with the fog and cloud node maintaining the energy consumption and execution delay balanced. For the comparison, we have used the function distribution as one of the matrices that indicate what percentage of functions in the serverless application are assigned to fog and cloud serverless platforms. This paper also provides a more in-depth standalone analysis of the DeF-DReL algorithm's performance in subsequent subsections. 

\subsection{Environment setup}\label{sec:perf:setup}
For the implementation of the DeF-DReL algorithm, which is basically a RL based algorithm, Tensorflow \footnote{https://www.tensorflow.org/} version 1.5.0 open-source machine learning platform with Python version 3.6.8 is used as the programming language through Jupyter Notebook version 4.4.0 \footnote{https://jupyter.org/}. Keras version 2.3.0 API is used atop Tensorflow framework. For the implementation of the proposed DeF-DReL algorithm, the total number of serverless applications ranges from $4$ through $10$ composing of $10$ serverless functions in each application. The size of a serverless function ranges from $10$ MB through $500$ MB, whereas the input size ranges from $100$ MB through $2500$ MB. Summarizing the resource requirement of serverless functions, the CPU, memory (RAM), storage, and network I/O resource requirements range from between $1-4$, $100 MB - 2048 MB$, $10 MB - 2048 MB$ and $10 KBps - 4096 Kbps$, respectively. It is to be noted that the critical value of the serverless function in the simulation is randomly distributed between the value $1$ and $5$. 

To align with the system model, there are two computing environments: \textit{Fog} and \textit{Cloud} computing environment. Both the environments have their computing, storage and network capabilities. With the available resource, the serverless platform possesses various limits, such as CPU limit, memory/RAM limits, etc. The list of parameters on resource limit, availability, requirements is given in Table \ref{table:simParam}.

\begin{table}[ht]
\caption{List of Parameters} \centering
\begin{tabular}{|p{0.3\linewidth}|p{0.25\linewidth}|p{0.25\linewidth}|}
    \hline
    Parameters  &\multicolumn{2}{c|}{\textbf{Value in ranges}} \\ \cline{2-3}
                                &Min value  &Max Value          \\  \hline
    \multicolumn{3}{|c|}{\textbf{SSR Bucket}}   \\  \hline
    Serverless functions/SSR	&4	    &10     \\  \hline
    \# of SSRs	                &4	    &10     \\  \hline 
    \multicolumn{3}{|c|}{\textbf{SERVERLESS FUNCTION}}          \\  \hline
    Size	                    &10	    &500    \\  \hline
    Input size	                &100	&2500   \\  \hline
    RAM demand	                &100	&2048   \\  \hline
    CPU demand	                &1	    &4      \\  \hline
    Storage Demand	            &10	    &2048   \\  \hline
    NW I/O demand	            &10	    &4096   \\  \hline
    Critical value	            &1	    &5      \\  \hline
    \multicolumn{3}{|c|}{\textbf{COMPUTING ENVIRONMENT}}        \\  \hline
            &CLOUD Environment &Fog Environment                 \\  \hline
    Input Limit         & 2500      &1500   \\  \hline
    CPU Limit           &6          &2      \\  \hline
    RAM Limit           &5120       &1024   \\  \hline
    Storage Limit       &10240      &1024   \\  \hline
    NW I/O Limit        &10240      &2048   \\  \hline
\end{tabular}\label{table:simParam}
\end{table}

\subsection{Simulation Results}\label{sec:perf:result}
\begin{figure}[ht]
    \centering
    \begin{subfigure}{.40\textwidth}
      \centering
        \includegraphics[width=0.95\linewidth]{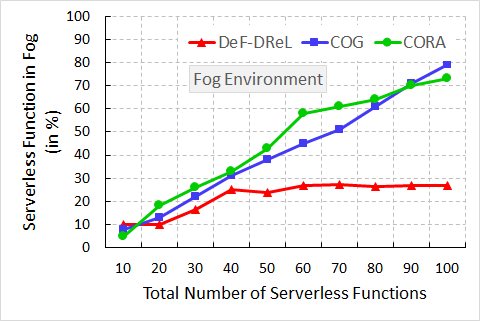}
        \caption{Percentage of serverless functions in fog.}
        \label{fig:perfEval:Fig1-FunDist_fog}
    \end{subfigure}
    \begin{subfigure}{.40\textwidth}
      \centering
        \includegraphics[width=0.95\linewidth]{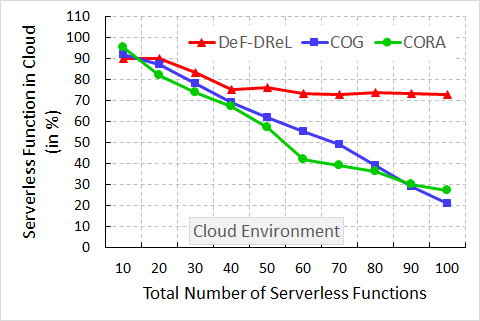}
        \caption{Percentage of serverless functions in cloud.}
        \label{fig:perfEval:Fig1-FunDist_cloud}
    \end{subfigure}
    \caption{Distribution of all serverless functions in fog and cloud.}
    \label{fig:perfEval:Fig1-FunDist}
\end{figure}

The performance in terms of the distribution of the serverless functions among fog and cloud serverless platforms of the proposed algorithm is evaluated by comparing with the performance COG\cite{8360511} and CORA\cite{8240666} algorithms. Figure \ref{fig:perfEval:Fig1-FunDist} shows the distribution of serverless functions indicating what percentage of functions is assigned to Fog serverless platform, as shown in Figure \ref{fig:perfEval:Fig1-FunDist_fog}, and cloud serverless platform, as shown in Figure \ref{fig:perfEval:Fig1-FunDist_cloud}, in case of the proposed and the other related algorithms. The X-axis and Y-axis in those figures represent the total number of serverless functions ranging from $10$ to $100$ and the percentage of serverless functions, respectively. It is to be noted that the size of each SSR ranges between $1-10$ number of serverless functions, and the number of SSRs is kept constant at $10$. The primary goal of DeF-DReL is to assign a minimum workload to fog as compared to the cloud environment without compromising the performance of the serverless application or violating the deadline of the user's application. On the other hand, the goals of COG and CORA algorithms is to assign more work to the fog instead of the cloud computing environment.

\begin{figure*}[ht]
    \centering
    \begin{subfigure}{.40\textwidth}
      \centering
        \includegraphics[width=0.95\linewidth]{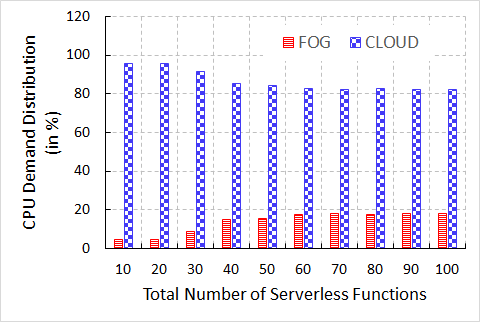}
        \caption{Distribution of CPU resource demand between fog and cloud.}
        \label{fig:perfEval:Fig2.1-ResrcDist_CPU}
    \end{subfigure}
    \begin{subfigure}{.40\textwidth}
      \centering
        \includegraphics[width=0.95\linewidth]{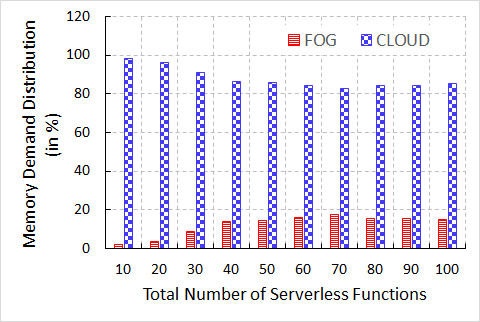}
        \caption{Distribution of memory resource demand between fog and cloud.}
        \label{fig:perfEval:Fig2.2-ResrcDist_Memory}
    \end{subfigure}
    \begin{subfigure}{.40\textwidth}
      \centering
        \includegraphics[width=0.95\linewidth]{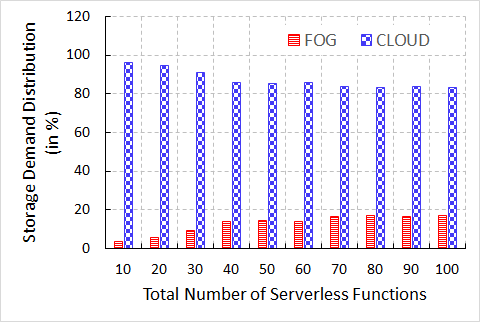}
        \caption{Distribution of storage resource demand between fog and cloud.}
        \label{fig:perfEval:Fig2.3-ResrcDist_Storage}
    \end{subfigure}
    \begin{subfigure}{.40\textwidth}
      \centering
       \includegraphics[width=0.95\linewidth]{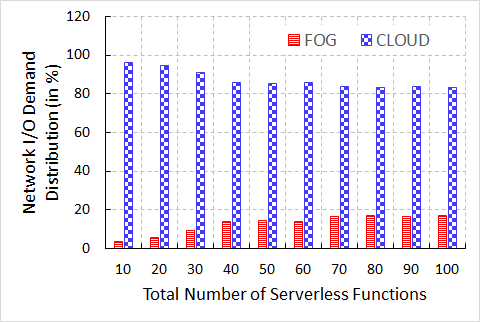}
        \caption{Distribution of Network I/O resource demand between fog and cloud.}
        \label{fig:perfEval:Fig2.4-ResrcDist_NW_IO}
    \end{subfigure}
    \caption{Distribution of different types of resource demand between fog and cloud.}
    \label{fig:perfEval:Fig2-ResrcDist}
\end{figure*}

With the DeF-DReL algorithm, the total percentage of the serverless functions that are assigned to fog serverless platform ranges from approximately $10\%$ to less than $30\%$, as in Figure \ref{fig:perfEval:Fig1-FunDist_fog}. Whereas approximately $70\% - 90\%$ of the functions are assigned to cloud environment, as shown in Figure \ref{fig:perfEval:Fig1-FunDist_cloud}. A very opposite trend is observed when COG\cite{8360511} and CORA\cite{8240666} algorithms are implemented. In the case of both the related algorithms, as the number of serverless functions increases, a steady growth in assigning functions to the fog environment is observed. Approximately $79\%$ and $72\%$ of the total $100$ serverless functions are assigned to fog when COG and CORA algorithms are implemented, respectively. Similarly, using COG algorithm, approximately $91\%$ of serverless functions are assigned to the cloud environment when a total of $10$ functions are distributed. This percentage declined to $21\%$ when the number of functions increases to $100$, as in Figure \ref{fig:perfEval:Fig1-FunDist_cloud}. A similar trend is also observed when the CORA algorithm is implemented, where approximately only $28\%$ of the total functions are assigned to the cloud and approximately $72\%$ of the total $100$ functions are assigned to the fog environment, as in Figure \ref{fig:perfEval:Fig1-FunDist_fog}. It is to be noted that no serverless functions are kept unassigned. This is due to the implementation design, where the learning agent needs to reach a state where all the functions are assigned to either fog or cloud computing environment. 

To further analyse the SSR workload distribution, we have investigated the resource demand distribution from the simulation result shown in, as shown in Figure \ref{fig:perfEval:Fig2-ResrcDist}. The implemented resource types are CPU, RAM or memory, storage and the Network I/O. The X-axis in all the sub-figures represents the total number of serverless functions, whereas the Y-axis represents the corresponding distribution of resource types in percentage. The CPU resource demand of all the functions allocated to the fog serverless platform is increasing from approximately $3\%$ to $18\%$ when the total number of functions increases from $10$ to $60$. However, the percentage of memory resource demand fulfilled by the fog environment remains less than $20\%$ even when the number of serverless functions is increased to $100$, as shown in Figure \ref{fig:perfEval:Fig2.1-ResrcDist_CPU}. A similar pattern is also observed in the other types of resources. Approximately $14\%$ and $86\%$ of RAM resources are fulfilled by the fog and cloud environments, respectively, when a total of $100$ serverless functions are deployed on both the computing environments, as shown in Figure \ref{fig:perfEval:Fig2.2-ResrcDist_Memory}. With the same number of serverless functions, i.e. $100$ functions, $16\%$ and $13\%$ of total storage resource demand and the network resource demands are fulfilled by the fog environment, as shown in Figure \ref{fig:perfEval:Fig2.3-ResrcDist_Storage} and \ref{fig:perfEval:Fig2.4-ResrcDist_NW_IO}, respectively. In a nutshell, it is observed that approximately $20\%$ of the total resource demands (including all types of resources) are fulfilled by the fog environment. This indicates that a larger number of serverless functions are mainly offloaded to the cloud to keep the fog environment available enough to accommodate more number of serverless functions from different users. One of the reasons behind such resource demand distribution is due to the limitations imposed on the serverless platform available in the fog environment, as shown in Table \ref{table:simParam}. This is mainly due to the fact that the fog environments are setup with limited resources with the intention to provide the necessary services to a maximum number of users.

\begin{figure*}[ht]
    \centering
    \begin{subfigure}{.4\linewidth}
      \centering
        \includegraphics[width=0.95\linewidth]{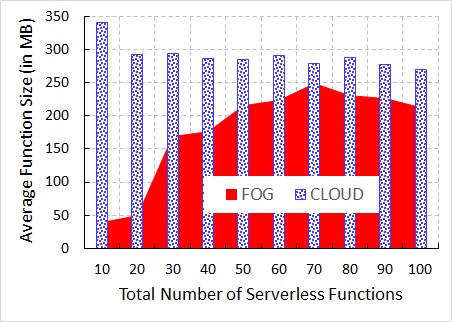}
        \caption{Average size of the functions.}
        \label{fig:perfEval:Fig3.0_funChar_AvgFunSize}
    \end{subfigure}
    \begin{subfigure}{.4\textwidth}
      \centering
        \includegraphics[width=0.95\linewidth]{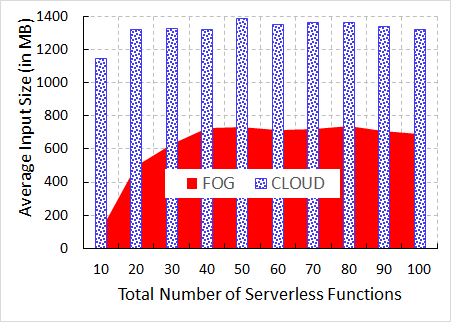}
        \caption{Average input size to the functions.}
        \label{fig:perfEval:Fig3.0_funChar_AvgFunInputSize}
    \end{subfigure}
    \caption{Characteristics (function size and input size) of serverless functions in fog and cloud.}
    \label{fig:perfEval:Fig3-FunChar}
\end{figure*}

\begin{figure*}[ht]
\centering
    \begin{subfigure}{.40\textwidth}
      \centering
        \includegraphics[width=0.95\linewidth]{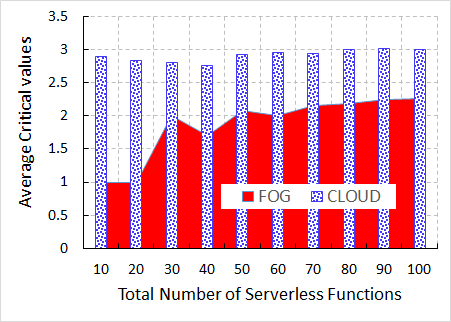}
        \caption{Average critical values of the functions.}
        \label{fig:perfEval:Fig3.0_funChar_AvgCriticalVal}
    \end{subfigure}
    \begin{subfigure}{.50\textwidth}
      \centering
       \includegraphics[width=0.95\linewidth]{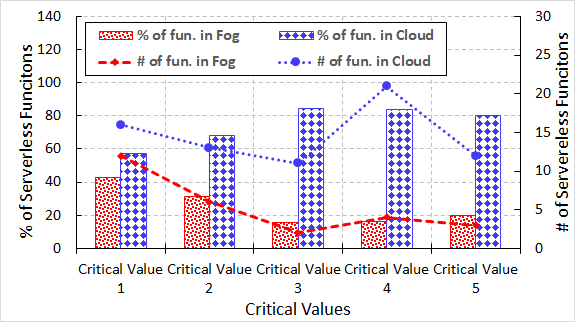}
        \caption{\# of functions with critical value in fog and cloud.}
        \label{fig:perfEval:Fig3.2_funChar_Critical_2}
    \end{subfigure}
    \begin{subfigure}{.7\textwidth}
        \includegraphics[width=0.75\linewidth]{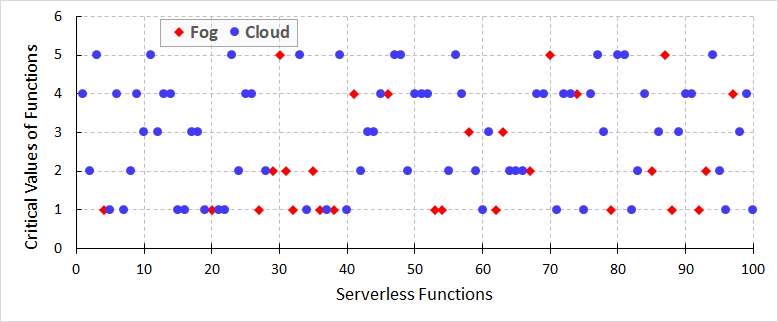}
        \caption{Critical values of the serverless functions in fog and cloud.}
        \label{fig:perfEval:Fig3.2_funChar_Critical}
    \end{subfigure}
    \caption{Characteristics (critical values) of Serverless functions in fog and cloud.}
    \label{fig:perfEval:Fig3-FunChar_critical}
\end{figure*}

\begin{figure*}[ht]
    \begin{subfigure}{.30\textwidth}
      \centering
        \includegraphics[width=0.95\linewidth]{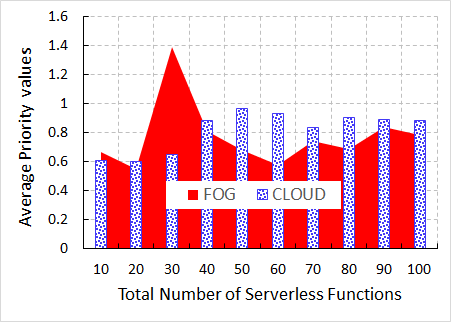}
        \caption{Average priority values of the functions.}
        \label{fig:perfEval:Fig3.0_funChar_AvgPriorityVal}
    \end{subfigure}
    \begin{subfigure}{.60\textwidth}
      \centering
        \includegraphics[width=0.95\linewidth]{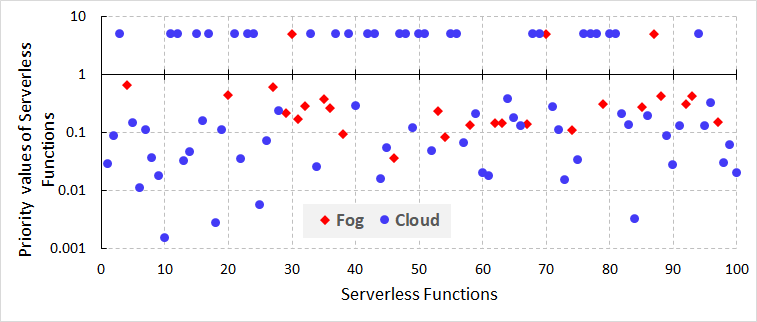}
        \caption{priority values of the serverless functions in fog and cloud.}
        \label{fig:perfEval:Fig3.1_funChar_Priority}
    \end{subfigure}
    \caption{Characteristics (priority values) of Serverless functions in fog and cloud.}
    \label{fig:perfEval:Fig3-FunChar_priority}
\end{figure*}

As discussed above that most of the resource demands are mainly fulfilled by the cloud environments, which allows the service providers to carefully use the fog resources. However, analysing the characteristics of the deployed serverless functions in both computing environments is an essential dimension to investigate. Accordingly, the Figure \ref{fig:perfEval:Fig3-FunChar}, \ref{fig:perfEval:Fig3-FunChar_critical}, and \ref{fig:perfEval:Fig3-FunChar_priority}, we have observed the size, input size, critical values and priority of the functions that are deployed in FN and in cloud environment. The sub-figures \ref{fig:perfEval:Fig3.0_funChar_AvgFunSize} and \ref{fig:perfEval:Fig3.0_funChar_AvgFunInputSize} gives an overall picture of the size and the input size of the functions. The size of the function indicates the total size of the serverless function code, which is in MegaByte (MB). The range of function code ranges from $10 MB$ to $500 MB$. Figure \ref{fig:perfEval:Fig3.0_funChar_AvgFunSize} shows that the average function size deployed in the fog environment is $250 MB$, which is less than that in the cloud environment. This average function size (deployed in FN) increases from less than $50~MB$ to $250~MB$ when the number of serverless function increases from $10$ to $70$. However, the value further decreases to approximately $210~MB$ when the number of functions further increased to $100$. On the other hand, it is observed that all the functions deployed in the cloud have a larger size ranging between $270~MB$ and $350~MB$. A similar trend observed while investigating the input size of those functions. The average size of the input to the functions in FN lies below $750~MB$, whereas it is larger than $1150~MB$ in the case of the functions deployed in cloud computing. This confirms that the algorithm is distributing the function as per requirement, keeping the load on the fog serverless platform minimal and offloading larger functions to the cloud.

To extend the simulation results' analysis and make sure that Def-DReL does not violate the constraints mentioned in Section \ref{sec:objFun}, we have presented the critical values of the functions deployed in fog and cloud environments, as in Figure \ref{fig:perfEval:Fig3-FunChar_critical}. Figure \ref{fig:perfEval:Fig3.0_funChar_AvgCriticalVal} shows the average critical value of the serverless function both in the fig and in cloud environments. The average critical values of the functions in a specific serverless platform are calculated by calculating the ratio of the sum of critical values of all the functions and the total number of serverless functions deployed in that serverless platform. From the figure, it is concluded that in the case of the fog serverless platform, the average critical values of the functions is less than the functions in the cloud environment. To complement this result, we have also analysed the total number of functions and percentage of all functions deployed in fog and cloud, as in Figure \ref{fig:perfEval:Fig3.2_funChar_Critical_2}. It is observed (also from Figure \ref{fig:perfEval:Fig3.2_funChar_Critical}) that a less number of functions are in the fog and a larger percentage of the functions are deployed in the cloud environment. For instance, there are a total of $15$ functions (out of $100$) are having the critical value $5$, out of which $3$ functions (accounted for $20\%$) are deployed in fog environment and the rest $12$ functions are deployed in the cloud environment. Similarly, out of $100$ serverless functions, a total of $28$ functions have the critical values $1$. Out of those $28$ functions, $12$ functions (i.e. $42\%$) are deployed in fog and $16$ functions (i.e. $57\%$) are deployed in cloud. Limit on the resource availability and the limit imposed by the fog serverless platform are some of the main reasons behind deploying higher critical valued functions on the cloud.

Based on the function size, input size and the critical values, the priority value of a serverless function is calculated, as defined in Equation \ref{eq:priority_fun}. To analyse the priorities of the serverless functions, we have depicted the simulation result in Figure \ref{fig:perfEval:Fig3-FunChar_priority}. Figure \ref{fig:perfEval:Fig3.0_funChar_AvgPriorityVal} represents the average priority of the serverless functions deployed in fog and cloud serverless platforms. The X-axis represents the average priority value and the Y-axis represents the total number of serverless functions. It is observed that when the number of functions increases beyond $40$, the average priority of the functions in fog become less than the functions in the cloud. This is due to the larger number of serverless functions are in the cloud affecting the final average value. In the case of $100$ functions, it is found that the average priority of the functions in fog ($27$ functions) is $0.77$, whereas the average priority value of functions in the cloud ($73$ functions) is $0.88$. From Figure \ref{fig:perfEval:Fig3.1_funChar_Priority}, it can be observed that most of the functions that are having the higher priority values are deployed in fog environment. It is to be noted that within an SSR, the functions with maximum resource demand or the highest critical value is assigned with the priority value $5$ instead of the value calculated using Equation \ref{eq:priority_fun}.

\section{Conclusions and future works}\label{sec:conc}
Efficient deployment of serverless applications, composed of several serverless functions, on fog and cloud computing environments is investigated in this paper. The underlined intention of introducing a fog computing environment is to provide services to the maximum number of connected users without compromising the service quality. To achieve this, this paper took advantage of offloading the optimum percentage of the user's request to cloud computing. This paper took advantage of deep reinforcement learning and proposed DeF-DReL, a systematic serverless functions deployment strategy for fog and cloud environments. The DeF-DReL strategy uses several parameters, such as distance and latency of the users from nearby fog node, user's priority in accessing the fog service, priority of the fog services, and resource demand of the user's request, to make it more realistic and applicable to a real-life scenario. The performance of the proposed deployment strategy is compared with the other existing strategies to observe its superiority and ability to achieve the concerned goal.

To further make it more realistic and applicable to the real-life scenario, it is necessary to incorporate a multi-fog and multi-cloud scenario, where a user is connected to multiple fog nodes and each fog node is connected to other multiple fog nodes, which is a part of our future works. Besides, to understand the real behaviour and performance, the proposed DeF-DReL deployment strategy will be implemented in a controlled lab environment in future.

\section*{Acknowledgment}
We acknowledge financial support to UoH-IoE by MHRD, India (F11/9/2019-U3(A)).

\bibliographystyle{elsarticle-num}
\bibliography{references}

\par \noindent
\begin{wrapfigure}{i}{1in}
\includegraphics[width=1in,height=1.25in,clip,keepaspectratio]{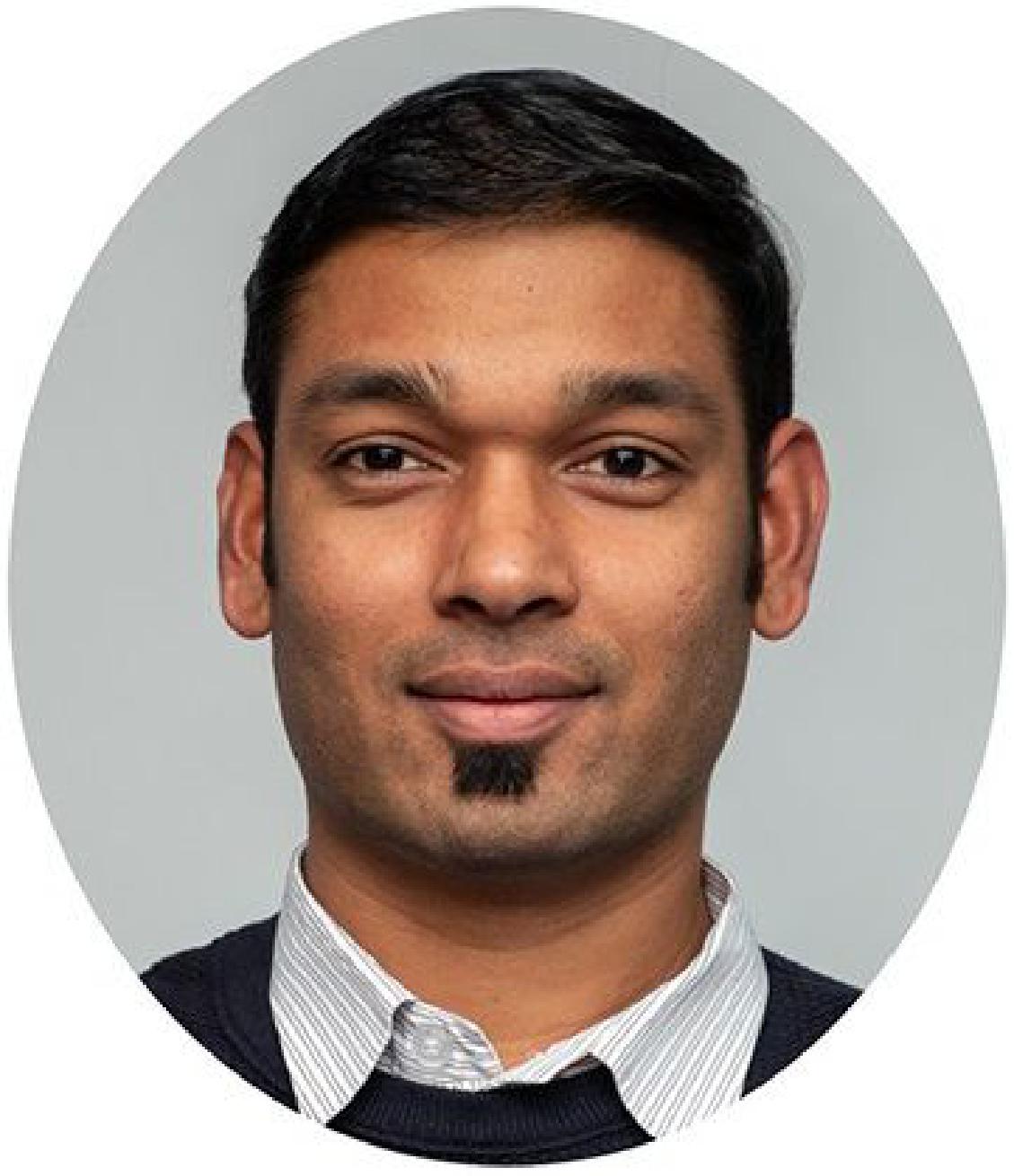}
\end{wrapfigure}
\textbf{Chinmaya Kumar Dehury}	received bachelor degree from Sambalpur University, India, in June 2009 and MCA degree from Biju Pattnaik University of Technology, India, in June 2013. He received the PhD Degree in the department of Computer Science and Information Engineering, Chang Gung University, Taiwan. Currently, he is a postdoctoral research fellow in the Mobile \& Cloud Lab, Institute of Computer Science, University of Tartu, Estonia. His research interests include scheduling, resource management and fault tolerance problems of Cloud and fog Computing, and the application of artificial intelligence in cloud management. He is an reviewer to several journals and conferences, such as IEEE TPDS, IEEE JSAC, Wiley Software: Practice and Experience, etc. 

\par \noindent
\begin{wrapfigure}{i}{1in}\includegraphics[width=1in,height=1.25in,clip,keepaspectratio]{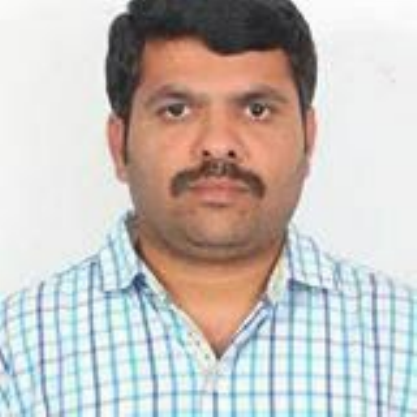}
\end{wrapfigure}
\textbf{Shivananda R Poojara} is a Ph.D student at Mobile and Cloud Computing Laboratory, University of Tartu, Estonia. He is also working as a Junior Research Fellow in IT Academy program. His research interests are serverless computing, edge analytics, fog and cloud computing. He is a member of IEEE, IET and ISTE. He has a 6+ years of teaching experience and worked at Nokia R\&D Labs as an intern.


\par \noindent
\begin{wrapfigure}{i}{1in}\includegraphics[width=1in,height=1.25in,clip,keepaspectratio]{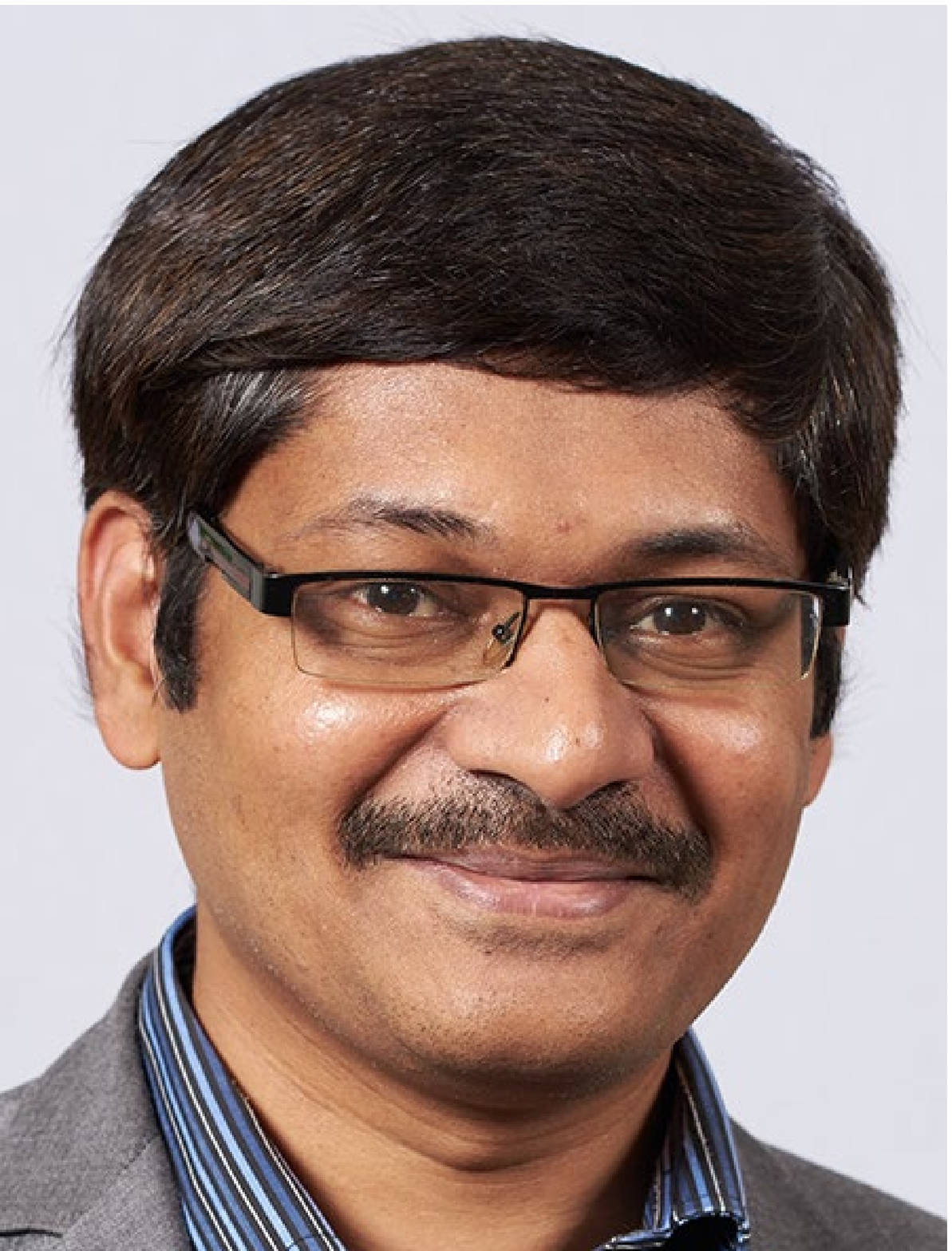}
\end{wrapfigure}
\textbf{Satish Narayana Srirama} is an Associate Professor at the School of Computer and Information Sciences, University of Hyderabad, India. He is also a Visiting Professor and the honorary head of the Mobile \& Cloud Lab at the Institute of Computer Science, University of Tartu, Estonia, which he led as a Research Professor until June 2020. He received his PhD in computer science from RWTH Aachen University, Germany in 2008. His current research focuses on cloud computing, mobile web services, mobile cloud, Internet of Things, fog computing, migrating scientific computing and enterprise applications to the cloud and large-scale data analytics on the cloud. He is an IEEE Senior Member, an Editor of Wiley Software: Practice and Experience, a 51 year old Journal, was an Associate Editor of IEEE Transactions in Cloud Computing and a program committee member of several international conferences and workshops. Dr. Srirama has co-authored over 165 refereed scientific publications in international conferences and journals. 

\end{document}